**Reflection on Purpose Changes Students' Academic Interests:**

**A Scalable Intervention in an Online Course Catalog**

Youjie Chen[a], Pranathi Iyer[b], René F. Kizilcec[a]

[a]Department of Information Science, Cornell University, Ithaca, NY, 14850, United States

[b]Computational Social Sciences, University of Chicago, Chicago, IL, 60637, United States

**Author Note**

Correspondence concerning this article should be addressed to Youjie Chen, Gates Hall 214, 107 Hoy Rd., Ithaca, NY 14853. email: yc2669@cornell.edu.

**Abstract**

College students routinely use online course catalogs to explore a variety of academic offerings. Course catalogs may therefore be an effective place to encourage reflection on academic choices and interests. To test this, we embedded a psychological intervention in an online course catalog to encourage students to reflect on their purpose during course exploration. Results of a randomized field experiment with over 4,000 students at a large U.S. university show that a purpose intervention increased students' cognitive engagement in describing their interests, but reduced search activities. Students became more interested in courses related to creative arts and social change, but less in computer and data science. The findings demonstrate the malleability of students' interests during course exploration and suggest practical strategies to support purpose reflection and guide students toward deliberate exploration of their interests in higher education.

**Introduction**

Interest exploration is a critical process that college students undertake (Flum & Kaplan, 2006). It expands students' intellectual horizons and helps them identify specific areas for subsequent interest development. Students who can identify and develop an interest are shown to attend classes more frequently, process instructional content more effectively, and perform better academically (Ainley et al., 2002; Hidi & Harackiewicz, 2000; Hidi & Renninger, 2006). Furthermore, interest exploration is beneficial beyond its immediate developmental effect: It plays an increasingly important role in students' preparation for future career development (Jordaan, 1963; Super, 1980). Today's students are expected to face future work environments with increasing numbers of career choices and career transitions due to rapid technological advancements, such as large language models, shaping new occupational structures (Eloundou et al., 2023). A willingness to engage in interest exploration is an important facilitator to adapt to the future of work (Flum & Blustein, 2000). Thus, supporting deliberate interest exploration in higher education can benefit students to not only achieve academic success but also be prepared for long-term career success.

In U.S. higher education, some institutions encourage interest exploration by offering students a broad range of course options (Stevens et al., 2018). Students look up course information using online course catalogs to facilitate course enrollment decisions. However, these platforms are not explicitly designed to promote interest exploration and interest-driven course selection. Even though online course catalogs store a complete set of course offerings by its institution, students only view a small fraction of course information based on what they search for. A recent study revealed that students only consider less than 2% of university courses listed on an online course catalog (Chaturapruek et al., 2021). Despite the prevalence of online course catalogs in universities, there is a limited understanding of how they shape students' access to course information and how they may be better used as a medium to support interest exploration during course selection.

Interest exploration during course selection is a complex process that involves identity and purpose exploration (Malin et al., 2014). Students are compelled to not only ask themselves questions like "What do I want to learn?" but more fundamentally, "Why do I want to learn this?" (Erikson, 1968; Pfund et al., 2020). Reflection on "why" an experience promotes students to engage in exploratory activities (Flum & Blustein, 2000). In particular, those who seek purpose through proactive engagement, as opposed to reaction to severe life events, have greater agency and openness to new experiences (Hill, Sumner, Burrow, 2014). Therefore, this research investigates purpose reflection as a catalyst to support interest exploration during course selection. Most prior work that has effectively facilitated purpose reflection among college students required elaborated programs where students participate in face-to-face discussion sessions on purpose (e.g., Bundick, 2011; Pizzolato et al., 2011; Thompson & Feldman, 2010; but for exceptions, see Reeves et al., 2021; Yeager et al., 2014). Online course catalogs, which are one of the most active and ubiquitous technologically-mediated contexts for academic decision making, have not been studied as a context for purpose reflection.

In the present study, we implemented a purpose reflection intervention by subtly integrating it into the search prompt of an online course catalog to facilitate interest exploration during course selection in higher education. We examined two types of purpose reflections, a self-interest-oriented purpose intervention and a self-transcendent-oriented purpose intervention, to understand how variations of a purpose intervention may influence students to explore their interests in different directions. The findings of this study advance our theoretical understanding of how student interest exploration can be influenced during course selection. It also demonstrates applied strategies for institution policymakers to encourage purpose reflection in an accessible, scalable, and equitable manner through the widely available technology of online course catalogs.

## Literature Review

### Online Course Catalogs

Almost every university has an online course catalog that stores official and basic course information to facilitate students' course exploration and selection. New advancements have been seen in online course catalogs in recent years. Some universities have begun to incorporate new types of information into online course catalogs, including course reviews, instructor ratings, student-reported course load, and even student course grades from previous years (e.g., University of Michigan, 2022; Stanford University, 2023). Other platforms utilize state-of-the-art recommendation algorithms to better match students with courses of interest (e.g., Pardos et al., 2019). These innovations have enhanced the utility of online course catalogs to help students navigate courses. They also highlighted the opportunity for researchers to examine the use of online course catalogs and how different types of information influence students' decision making.

Alongside the practical significance of online course catalogs, their methodological advantages for research have gained prominence. Traditionally, phenomena around academic decision making, such as course and major choice, are mainly studied through qualitative or survey methods (e.g., Babad & Tayeb, 2003; Goyette & Mullen, 2006; Harrison et al., 2022). Online course catalogs have historically not been used for research, only for administrative purposes. In recent years, however, the rise of educational data science has led online course catalogs to become valuable research instruments to collect large-scale authentic data and even run randomized field experiments to answer research questions in education (McFarland et al., 2021).

Compared to previous research methods, they can reveal empirical traces of how students consider course options before arriving at a final decision. For example, by analyzing clickstream data from a university-wide online course catalog, Chaturapruek and colleagues (2021) found that the number of courses that students consider is much more restricted than the number of courses offered by the university. Moreover, online course catalogs can incorporate randomized experiments to identify causal relationships during academic decision making with high statistical power and at a low cost. For example, Chaturapruek and colleagues (2018) ran a campus-wide randomized field experiment on an online course catalog to examine the effect of varied access to grade information and time commitment information on students' course choices. They found that changes in the type of information presented did not affect the composition of courses that students choose, but access to grade information surprisingly lowered students' overall GPAs. In another example, Yu and colleagues (2021) conducted a randomized controlled experiment on an online course catalog to investigate the effect of different course recommendation engines on course choices. The study revealed that a search engine was most effective in promoting serendipitous course discovery when it explains new course recommendations with keywords from courses that a student took before. In summary, these studies demonstrate that online course catalogs can provide new complementary perspectives to understand students' course exploration and selection process. Following this line of research, we conducted a large-scale randomized field experiment through an online course

INTEREST EXPLORATION IN AN ONLINE COURSE CATALOG                                        7catalog and analyzed variations in students' digital trace data to understand how students explore their interests.

**Interest Exploration and Purpose Reflection**

We define interest exploration based on the classic interest development model proposed by Hidi and Renninger (2006). Interest exploration is the stage where students explore a range of situational interests that are initially unstable but may potentially become long-lasting individual interests. We thus theorize two mechanisms through which interest exploration may be affected by purpose reflection.

First, purpose reflection may affect interest exploration by helping students develop a clearer understanding of their interests. College students commonly struggle to come up with ideas of what to study during course search and selection, which can prevent them from engaging in interest exploration (Chen et al., 2022). Proactive efforts to seek out purpose, such as explicit reflection on purpose, can bring clarity to their purpose and induce concrete thoughts on ways to pursue it (Kashdan & McKnight, 2009). For example, a student may identify that their purpose is to create a more sustainable living space for the next generation. This idea may prompt them to explore interests in environmental engineering or biology, which they may have never realized before reflecting on their purpose. Therefore, a purpose reflection intervention may give students new ideas on what interests to explore and promote students to engage more in interest exploration.

However, the process of reflecting on purpose is not always enjoyable. In fact, previous research shows that engaging in self-exploration of identity and purpose may cause psychological distress (Kidwell et al., 1995; Schwartz et al., 2009). As a result, students may want to avoid feelings of frustration and not engage in purpose reflection. This raises the possibility that a purpose reflection intervention may unintentionally discourage students from exploring their interests during course selection. Given the two possible outcomes, we raise the first research question:

**RQ1:** How does a purpose reflection intervention impact students' engagement with interest exploration?

A second way that purpose reflection may affect interest exploration is by signaling students' interest-driven motivation. Interest is only one of the many motivators that influence a student's course decision (Babad et al., 1999; Babad & Tayeb, 2003; Hinds & Shultz, 2018). Other main motivators include fulfilling course requirements, achieving high grades, and managing course workloads (Babcock, 2010; McGoldrick & Schuhmann, 2002; Ognjanovic et al., 2016). These motivators can sometimes be at odds with each other during course selection. Interest-driven choices may be salient or inconspicuous depending on which motivator is signaled when the individual enters a context (Sansone & Thoman, 2006). For example, a student may not engage in interest-driven course exploration if they are primed to be overly concerned about their grades. Under such circumstances, reflection on purpose may encourage students to prioritize interest as a factor in making course choices given that reflection on purpose can heighten intrinsic motivation (Littlejohn, 2008). In particular, having a purpose often leads people to engage in activities they find more personally meaningful (Nakamura & Csikszentmihalyi, 2002; Renninger, 2000). Relatedly, previous interview studies show that when high school students discuss future career interests, they spontaneously tie them to a purpose (Hulleman & Harackiewicz, 2009). Therefore, there is good reason to believe that reflection on purpose may influence students to explore their interests, particularly in areas they find personally meaningful.

What academic areas would an individual find personally meaningful? Here we distinguish areas of interest led by two types of purpose, a self-interest purpose and a self-transcendent purpose. This distinction comes from two streams of research on purpose in the literature. Damon and colleagues (2003) define purpose as "a stable and generalized intention to accomplish something that is at once meaningful to the self and of consequence to the world beyond the self." Embedded in this definition is the emphasis on a self-transcendent element in purpose. That is, a purpose needs to be tied to a desire to contribute to the world other than being of personal interest. Other scholars study purpose that is more broadly defined as a central aim that directs life goals (McKnight & Kashdan, 2009; Ryff, 1989), which can be of self-interest as well. Prior research shows that when adolescents talk about their purpose, a self-interest purpose and a self-transcendent purpose are distinguishable, but often coexist (Yeager et al., 2012; Yeager & Bundick, 2009). However, there is limited empirical evidence on whether the distinction between a self-transcendent purpose and a self-interest purpose is necessary among college students and whether they may lead to different directions of interest. In this study, we developed two interventions based on these two conceptualizations of purpose to examine if they would cause any difference in students' interest exploration. Specifically, we ask the following research question:

**RQ2:** How does reflection on different types of purpose (self-interest purpose and self-transcendent purpose) influence the areas of interest that students explore?

## The Present Study

We investigated the role of purpose reflection in interest exploration during course searches. We developed a low-cost, brief, and scalable approach to facilitate purpose reflection by reframing the course search prompt on an online course catalog that is in regular use by thousands of students at the institution. We examined two kinds of framing messages, one that emphasizes self-interest purpose and one that emphasizes self-transcendent purpose, and compared them against a control condition that did not prompt students to reflect on their purpose.

      We first examined how the two purpose reflection interventions may affect students to engage in interest exploration during course search (RQ1). Specifically, we examined the effects on both students' behavioral engagement and cognitive engagement. These two dimensions of engagement are conceptually different but related (Ben-Eliyahu et al., 2018). For example, a student may not show behavioral engagement in a classroom by not raising their hand, but they may engage cognitively by thinking about the questions that the teacher raised. Measuring both behavioral and cognitive engagement can help us better explore the specific mechanisms for a purpose reflection intervention to be effective (Archambault et al., 2009). Based on our literature review, we remain agnostic as to whether a purpose intervention will positively or negatively affect students' engagement in either dimension. It is also possible that the intervention does not have any effect on students' engagement at all given that the intervention is brief and subtle.

      We then examined how purpose reflection may affect the areas of interest that students explore (RQ2). We hypothesized that students who were given either a self-interest purpose intervention or a self-transcendent purpose intervention would be more interested in areas that can be associated with their own purpose compared to a control condition. Specifically, in the self-interest purpose intervention, students may explore areas of interest that support purpose towards the self, such as "to enjoy". In the self-transcendent purpose intervention, students may explore areas of interest that support their purpose towards others, such as "to help others" or "to protect the planet." To systematically categorize students' areas of interest and formulate

hypotheses, we apply topic modeling to students' course search queries across all three conditions (see Methods for details). This data-driven approach is agnostic to the areas students might be interested in. We identified five areas of interest: (1) business, (2) computer and data science, (3) creative arts, (4) social change, and (5) all other areas of interest.

Following the topic modeling results, we formulated hypotheses for the first four specific areas of interest. First, we hypothesized that a purpose intervention would cause students to be less interested in business and in computer and data science. This is grounded in prior literature on purpose, which suggests that purpose is generally tied to more intrinsic than extrinsic motives such as financial gain or status (Yeager et al., 2012; Yeager & Bundick, 2009). Even though areas of computer and data science and business can be intrinsically self-fulfilling and create benefits for others, they are often viewed as areas that lead to high-paying jobs (Binder et al., 2016). Second, we hypothesized that the self-interest purpose intervention would increase students' interest in creative arts, because self-interest purpose emphasizes intrinsic motives towards the self, such as studying an area that is personally enjoyable. The creative arts are often associated with play and enjoyment (An, 2019; Schmidhuber, 2010). Third, we hypothesized that the self-transcendent purpose intervention would increase students' interest in social change, because self-transcendent purpose emphasizes intrinsic motives towards others, such as studying in an area that benefits the community. These three hypotheses guide our examination of how academic interests may be affected by purpose reflection during course searches.

## Methods

### Context and Participants

The research is conducted using an online course catalog platform at a large and selective land-grant university in the United States. On the platform's landing page (Figure 1), a search prompt is displayed: "*Discover Cornell courses and Explore Your Interests. Describe your ideal course in a few sentences: What do you want to learn more about? What skills do you want to master? What learning format do you like best?*" A large textbox is placed underneath the search prompt to provide students ample space to articulate their interests. Upon submitting their search query, they see the search result page (Figure 2), which shows a list of courses relevant to the search query. Courses are presented with the course title and description, course code, instructor names, grading basis, and any prior offerings. The list of courses is ranked by the cosine similarity between the search query and the course title and description, which is a standard algorithm used by many information retrieval platforms.

**Figure 1**

*Screenshot of the Landing Page of the Online Course Catalog in the Control Condition.*

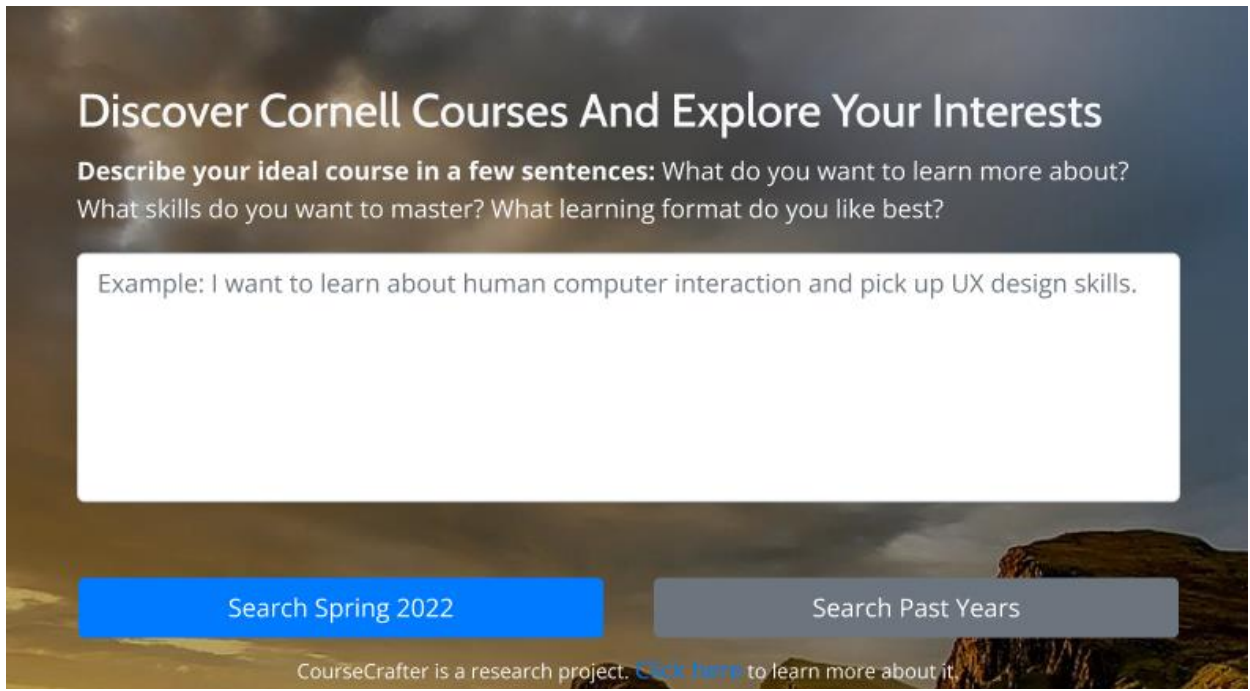

**Figure 2**

*Screenshot of the Search Result Page of the Online Course Catalog*

  The online course catalog is a campus-wide resource that anyone at the university can have access to since its official launch in February 2020. Participants' log data was merged with student demographic data obtained from the university registrar using hashed student IDs. Most participants (91.3%; N = 4,032) are students for whom we have demographic records. The remaining participants are faculty and staff at the institution. The analysis was only conducted for student participants. Among them, most are undergraduates (84.2%), and some are graduate (7.8%) and professional students (8.0%). The demographic make-up is as follows: 55.3% female, 44.7% male; 31.0% White, 18.4% Asian, 12.7% Hispanic, 6.6% Black, 4.8% mixed-race, and 26.4% other or unknown race.

  The online course search tool is used especially 1-2 weeks leading up to the course enrollment deadline. The timeframe of the present study extended from the start of Fall '21 to the end of Spring '22. During this period students encountered multiple course pre-enrollment and add/drop periods. In total, the platform collected 7,500 search queries from participants.

**Procedure**

Upon first arrival at the landing page, participants were randomly assigned to one of three conditions with equal probability: the self-interest purpose intervention (N = 1,475), the self-transcendent purpose intervention (N = 1,460), and the control condition (N = 1,483). Participants were unaware of their condition assignment and remained in the same condition throughout the study period. Based on the condition, we manipulated two static components of the landing page: the prompt above the search textbox and the example search query in the search textbox (see Table 1, for the exact text; Figure 1 shows the control condition). The manipulated text was shown for every search during the study period. We crafted the search prompt and search example texts based on prior research by Yeager et al., (2012), who identify distinguishing features between a self-interest and a self-transcendent purpose. In each condition, one out of three example search queries was selected at random to be displayed on the search page. In the control condition, the example queries followed the format of "I want to learn about…" to describe an interest without describing a purpose. In the treatment conditions, the example queries followed the format of "I want to learn about… because…" to describe an interest with a purpose. Specifically, the example in the self-interest purpose condition focuses on purposes in relation to one's skills or desires. The example in the self-transcendent purpose condition focuses on purposes in relation to making a positive impact to something larger than the self. The example search query was created as a worked example to help participants understand and follow the search prompt. Except for the search prompt and the example query, the interface and functionality are identical across all three conditions.

**Table 1**

*Search Prompt and Example Search Query in Each Condition*

| Manipulations | Self-Interest Purpose Condition | Self-Transcendent Purpose Condition | Control Condition |
|---|---|---|---|
| Search Prompt | Discover Cornell Courses and Pursue Your Dreams. Describe your ideal courses in a few sentences: What do you want to learn more about and why? What skills do you want to master and why is it important to you? | Discover Cornell Courses and Make the World a Better Place. Describe your ideal courses in a few sentences: What do you want to learn more about and why? What skills do you want to master and why is it important to you? | Discover Cornell courses and Explore Your Interests. Describe your ideal course in a few sentences: What do you want to learn more about? What skills do you want to master? What learning format do you like best? |
| Search Example (one is displayed at random) | Example: I want to learn about digital art because I like to be an artist and work in the creative industry. | Example: I want to learn about sustainability because I think it addresses a lot of environmental concerns. | Example: I want to learn about human-computer interaction and pick up UX design skills. |
|  | Example: I want to learn about artificial intelligence because it will help me find jobs that pay well. | Example: I want to learn about artificial intelligence because I believe it can have a positive impact on people's lives. | Example: I want to learn about how artificial intelligence is used in real estate. |
|  | Example: I want to learn about quantum physics because I think it advances our knowledge of the universe. | Example: I want to learn about public health because I want to find better ways to respond to a global crisis like covid. | Example: I want to learn about the theory of dreams, Freud, and psychoanalysis. |

**Measures**

*Engagement in Interest Exploration*

We measured two dimensions of engagement in interest exploration: behavioral engagement and cognitive engagement. Behavioral engagement is measured by students' search activities on the platform using two indicators. The first indicator is whether a student conducted at least one search on the platform during their initial use. The second indicator is whether a student returned to the platform another time after their initial use.

Cognitive engagement is measured based on how students engage in writing search queries. This measure is limited to students who conducted at least one search. We operationalized cognitive engagement using two indicators. The first indicator is the average number of words in a student's search query. Prior work has shown that the number of written words can provide a valid measure of higher levels of cognitive processing (Joksimovic et al., 2014; Tausczik & Pennebaker, 2010). The second indicator is whether at least one search query from the participant showed evidence of causal thinking. Prior work suggests that purpose is often seen as addressing the "why" question behind an interest (Yeager & Bundick, 2009). We used the causation index in the LIWC22 dictionary to identify if a search query contains words that indicate causal statements (Boyd et al., 2022). To reduce the number of false positives due to keywords related to causation (e.g., the search query "I'm interested in causal inference" has causal keywords but is not a causal statement), we coded all search queries with less than six words (73.2% of all queries) as not presenting causal thinking. We confirmed that there is no significant difference in the number of queries with less than six words across conditions ($F(2, 2090) = 2.33$, $p = 0.10$).

*Areas of Interest*

We measure areas of interest using a topic model fitted to all search queries and an individual-level probability distribution over the derived topics. We used search queries to derive areas of interest because their open-ended nature can provide an authentic representation of the areas that students are interested in, regardless of the performance of the search algorithm or whether the university offers courses in those areas. The topic model (details in Analytical Approach) identified five areas of interest: business, computer and data science, creative arts, social change, and all other topics. Based on all search queries a student submitted during the experiment period, we obtain a vector of probabilities over the five topics to characterize their areas of interest (e.g., for a student, their area of interest may be 70% business, 10% computer and data science, 3% create arts, 2% social change, 15% other).

**Analytical Approach**

*Regression Analysis*

We used logistic regression models to analyze the intervention effect when the dependent variable is binary (e.g., whether a student returned to the site, or whether a student searched). We used linear regression models with robust (i.e. heteroskedasticity-consistent) standard errors to estimate treatment effects for the two continuous outcomes, which are the average query length and the topic probability of each area of interest in search queries derived from topic modeling results.

We separately analyzed outcomes for new and existing users of the search platform. New users are students who started using the platform after the experiment launched (61.6%), while existing users have used it at least once before the experiment (38.4%). This distinction is important because the text-based manipulation of our intervention is more likely to be overlooked by existing users who are familiar with the platform. We expect that the purpose intervention will have a stronger effect on new users than on existing users.

*Topic Modeling*

We employed topic modeling to find the areas of interest that participants indicated in their search queries (for this, we concatenate all of their queries into one text). Topic modeling is an unsupervised machine learning approach that identifies latent topics in documents based on the recurrence and co-occurrence of words. We used latent Dirichlet Allocation (LDA; Blei et al., 2003) to perform topic modeling. LDA represents each document as a mixture of latent topics with topic probabilities and represents each topic as a mixture of words with word probabilities. We describe the pre-processing steps and topic identification process in more detail.

**Pre-Processing.** To prepare the inputs for the LDA algorithm, we first concatenated search queries across all semesters during the experiment period to represent a "document" for each student. This is because short documents, such as Twitter posts, provide worse topic modeling results due to their limited information value (Hong & Davison, 2010). Concatenating student queries results in longer documents that can substantially increase the quality of results (Zhao et al., 2011). Also, it is likely that a search query only represents one topic instead of a mixture of topics, threatening the basis of the LDA model (Blei, 2012). Concatenating queries on the participant level allows us to understand the mixture of topics that a participant explored across all searches. Second, we tokenized each document into both unigrams and common bigrams. Third, we removed common stop words using the predefined list in the Natural Language Toolkit (NLTK) and additional ones unique to our research context (e.g. 'class',

'study'). These pre-processing steps improved the topic modeling results. For details of the pre-processing steps, see Appendix A.

**Identifying Areas of Interest.** We used the tokenized documents as inputs to perform topic modeling using the *Gensim* package version 4.2 (Rehurek and Sojka, 2011). LDA requires the number of topics to be specified as an input. Determining the optimal number of topics is typically an iterative decision process that relies on the researchers' discretion based on the research purpose (Chang et al., 2009). We systematically evaluated the topic modeling results across a range of topic numbers using the UMass metric to assess within-topic coherence (Röder et al., 2015). We also used the most frequent tokens and most representative documents in each topic to evaluate the qualitative interpretability of the topic model results (Jacobs & Tschötschel, 2019). For details of the evaluation process, see Appendix B.

In the end, the topic model identified eight topics. We then combined several related topics into five areas of interest to arrive at a comparable level of granularity across areas. The five areas of interest are (1) business (combined 2 topics), (2) computer and data science, (3) creative arts (combined 2 topics), (4) social change, and (5) others (combined 2 topics). Table 2 summarizes the areas of interest with the average probability of each area across all conditions, examples of the most frequent terms related to the area, and example queries in each area.

**Table 2**

*The Five Areas of Interest, the Average Probability of Each Area Across All Conditions, Examples of Most Frequent Terms Related to the Areas of Interest, and Example Queries*

| Areas of Interest | Average Probability across All Conditions | Examples of Most Frequent Terms | Example Query |
|---|---|---|---|
| Business | 22.0% | Business, law, management, marketing, consulting, finance, leadership | *"I want to learn about the business side of the biotechnology industry including venture capital and management"* |
| Computer and Data Science | 18.2% | Data, science, learning, machine, machine learning, computer, data science, theory, intelligence, design | *"I want to learn about virtual reality machine learning and computer science"* |
| Creative Arts | 22.9% | Art, arts, history, sciences, arts sciences, latin, design, research, writing, music, creative | *"I want to understand how to get talents to thinking creatively to give an organization a competitive advantage"* |
| Social Change | 12.1% | Climate, animals, world, change, science, climate change, law, politics, philosophy, social | *"I want to learn about feminism and misogyny across the world and how gender roles have changed over time"* |
| Others | 24.8% | Human, psychology, biology, health, art, engineering, science, medicine, quantum | – |

**Results**

**Effect on Engagement in Interest Exploration**

First, we examined students' behavioral engagement in terms of whether they conducted a search and whether they returned to the platform. We found that the self-interest and self-transcendent purpose interventions resulted in behavioral disengagement in terms of students' likelihood to search, especially for new users, but not in terms of their likelihood to return. Specifically, new users were significantly less likely to search with the self-interest purpose intervention (43.0%; logOR = -0.38, $z$ = -3.73, $p$ < 0.01) and self-transcendent purpose intervention (42.8%; logOR = -0.37, $z$ = -3.64, $p$ < 0.01) compared to the control condition (52.1%). Existing users were less affected: They were less likely to search with the self-transcendent purpose intervention (54.8%; logOR = -0.32, $z$ = -2.57, $p$ = 0.01) compared to the control condition (62.5%), but remained similar with the self-interest purpose intervention (64.1%, logOR = 0.07, $z$ = 0.58, $p$ = 0.56). Neither intervention significantly affected whether students would return to the platform, no matter if they were new or existing users (self-interest purpose intervention: for new users, logOR = -0.06, $z$ = -0.57, $p$ = 0.57, for existing users, logOR = 0.04, $z$ = 0.30, $p$ = 0.77; self-transcendent purpose intervention: for new users, logOR = -0.03, $z$ = -0.31, $p$ = 0.75, for existing users, logOR = 0.10, $z$ = 0.81, $p$ = 0.42).

Second, we examined students' cognitive engagement based on how they wrote their search queries (analyzing data for the 2,092 students who conducted at least one search). We found significant intervention effects on cognitive engagement in query length (Figure 3) and causal thinking (Figure 4). Specifically, both the self-interest and self-transcendent purpose intervention encouraged students to write longer search queries (self-interest purpose intervention: for new users, mean diff. = 2.94, $t(1097)$ = 5.15, $p$ < 0.01, for existing users, mean diff. in words = 2.14, $t(989) = 3.79$, $p < 0.01$; self-transcendent purpose intervention: for new users, mean diff. = 2.50, $t(1097) = 4.24$, $p < 0.01$, for existing users, mean diff. = 1.43, $t(989) = 2.61$, $p = 0.01$).

Moreover, both the self-interest and self-transcendent purpose intervention encouraged students to use more causal words in their search queries. In the self-interest purpose condition, the odds were 2.70 times greater for new users to write their search queries in a causal way and the odds for existing users were 1.86 times greater (logOR = 0.99, $z = 4.67$, $p < 0.01$; logOR = 0.62, $z = 2.78$, $p < 0.01$). In the self-transcendent purpose condition, the odds for new users were 2.29 times greater and the odds were 1.69 times greater for existing users to write their search queries in a causal way (logOR = 0.83, $z = 3.84$, $p < 0.01$; logOR = 0.53, $z = 2.23$, $p = 0.03$).

**Figure 3**

*The Effect of Purpose Reflection Interventions on Search Query Length*

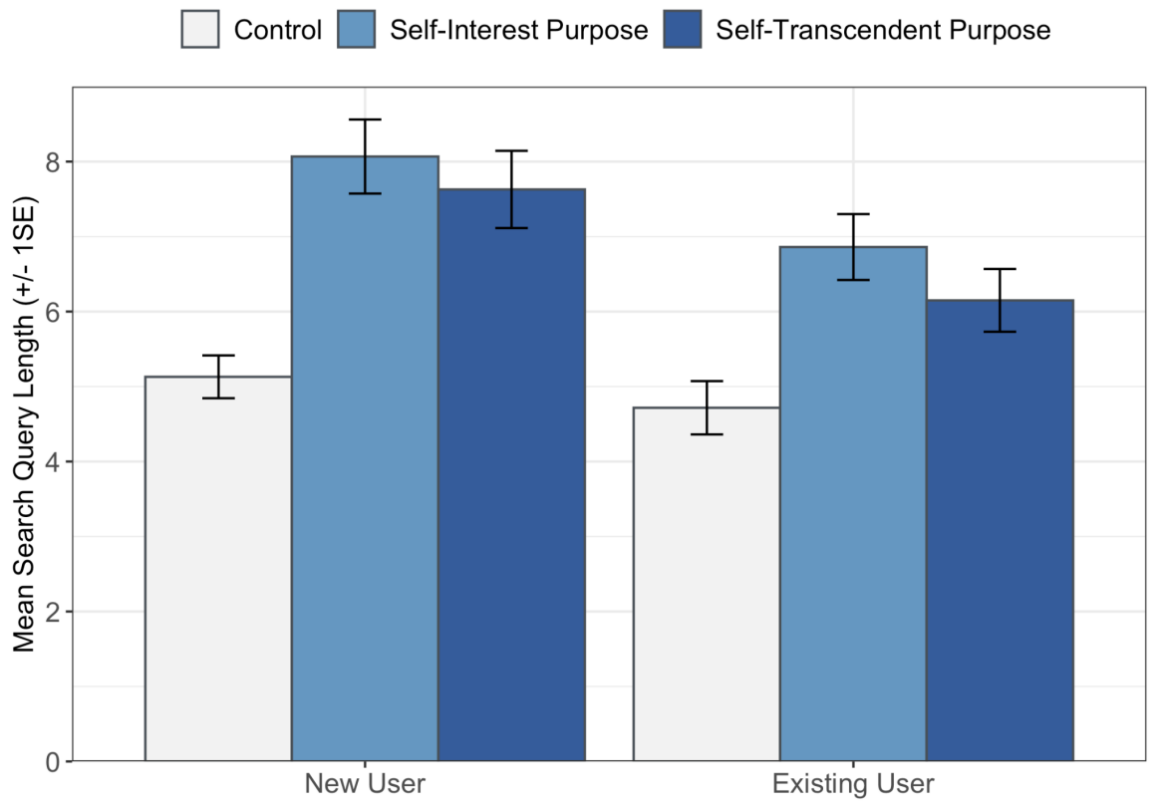

**Figure 4**

*The Effect of Purpose Reflection Interventions on the Percentage of Causal Thinking Displayed in Students' Search Queries*

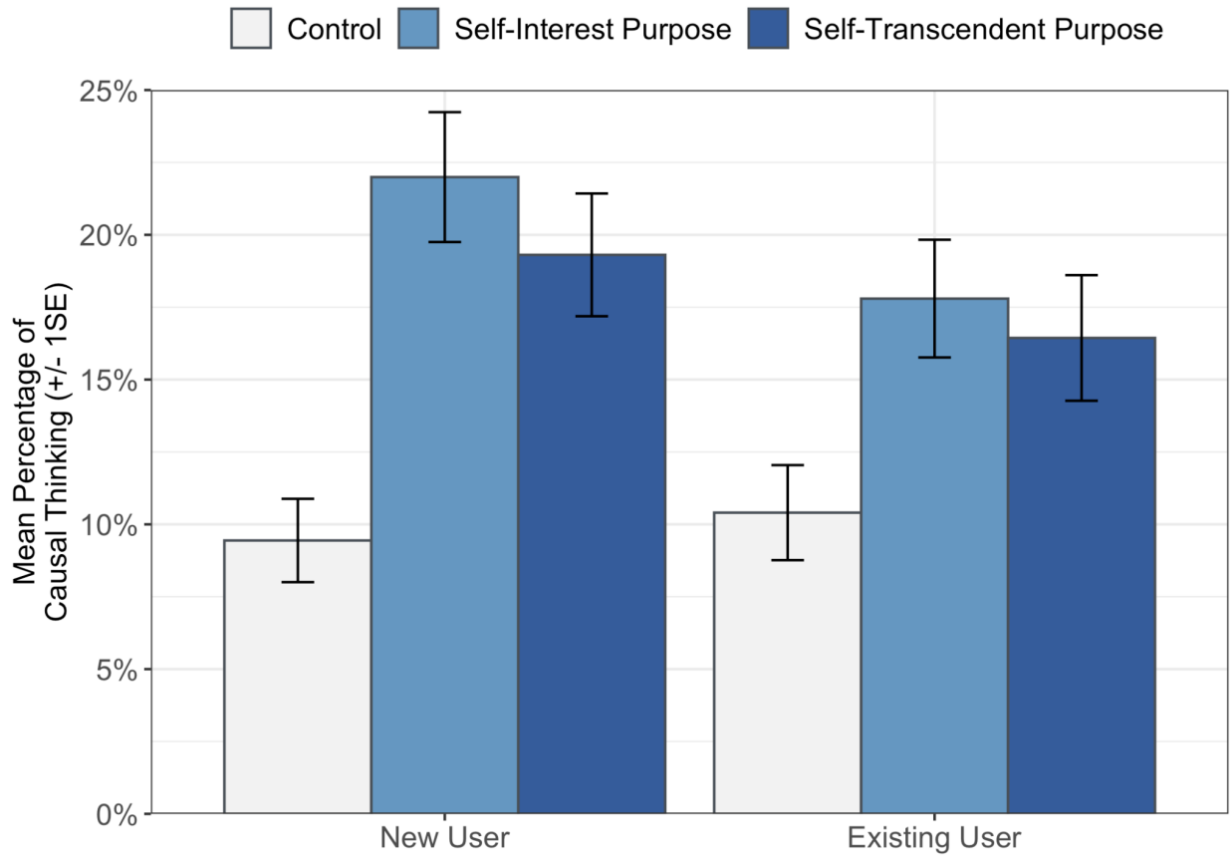

**Effect on Area of Interest**

We hypothesized that the self-interest purpose intervention would increase students' interest in creative arts, the self-transcendent purpose intervention will increase interest in social change, and both interventions will decrease interest in business and computer and data science. Figure 5 illustrates the difference in students' areas of interest across the three conditions separately for new and existing users. We found that both the self-interest purpose intervention and the self-transcendent purpose intervention caused new users to be more interested in creative arts and social change, but less interested in computer and data science. The intervention did not affect their interest in business. For existing users, only the self-transcendent intervention affected students to be more interested in social change. We report the specific intervention effects for each area of interest.

**Figure 5**

*The Effect of Purpose Reflection Interventions on Areas of Interest*

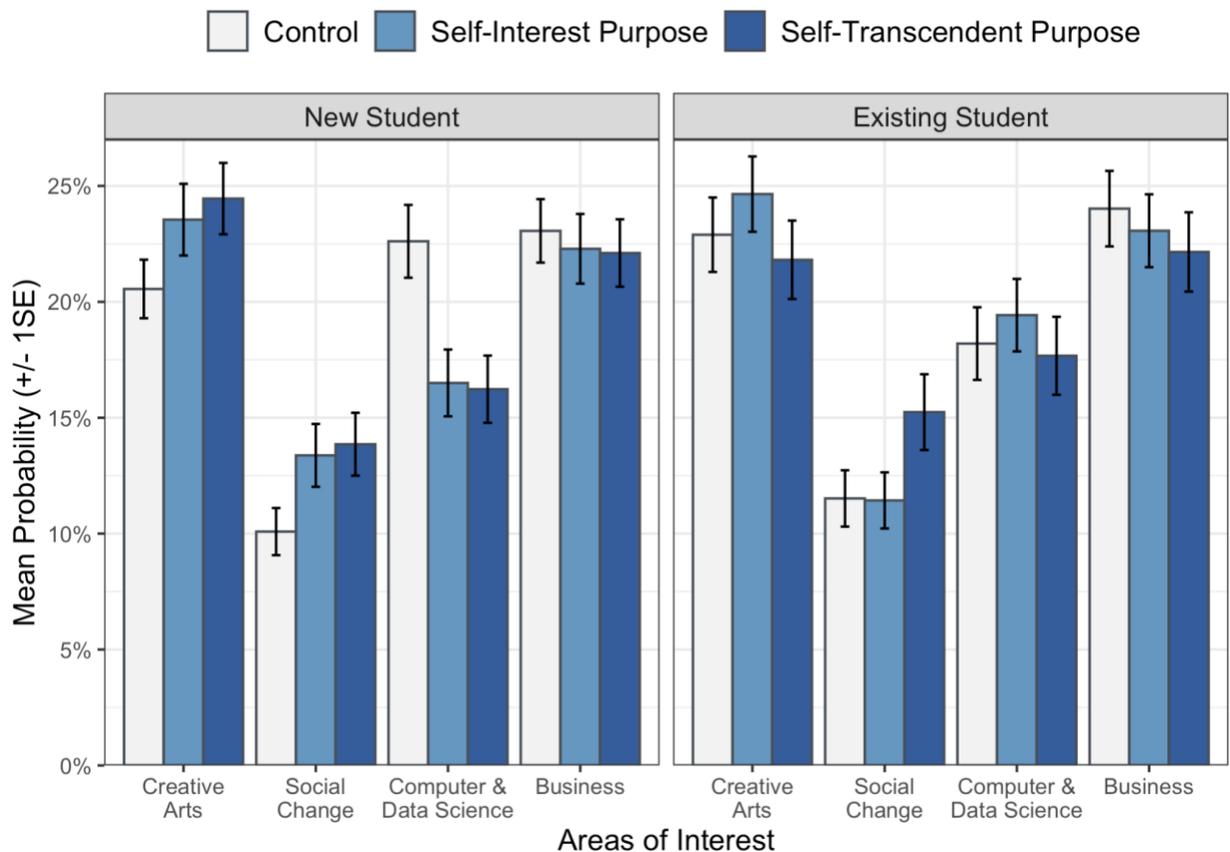

For new users, the self-transcendent purpose intervention increased their interest significantly in creative arts from 20.1% to 24.8% ($t(1097) = 2.19$, $p = 0.03$) and in social change from 10.1% to 13.9% ($t(1097) = 2.23$, $p = 0.03$). In contrast, it led students to be significantly less interested in computer and data science from 22.6% to 16.2% ($t(1097) = -2.99$, $p < 0.01$). It did not cause a significant change of interest in business (percentage point change = -0.8%, $t(1097) = -0.37$, $p = 0.71$). The self-interest purpose intervention caused a similar shift of interest in these areas of interest (creative arts: pp change = 3.7%, $t(1097) = 1.75$, $p = 0.08$; social change: pp change = 3.3%, $t(1097) = 1.94$, $p = 0.05$; computer and data science: pp change = -6.1%, $t(1097) = -2.87$, $p < 0.01$; business: pp change = -0.3%, $t(1097) = -0.15$, $p = 0.88$).

For existing users, the self-transcendent purpose intervention increased their interest in social change from 11.5% to 15.2% ($t(989) = 1.83$, $p = 0.07$), but it did not change their interest in other areas (creative arts: pp change = -1.6%, $t(989) = -0.68$, $p = 0.50$; computer and data science: pp change = -0.5%, $t(989) = -0.23$, $p = 0.82$; business: pp change = -1.8%, $t(989) = -0.73$, $p = 0.46$). The self-interest purpose intervention did not significantly change students' areas of interest (creative arts: pp change = 1.2%, $t(989) = 0.52$, $p = 0.60$; social change: pp change = 0.0%, $t(989) = 0.02$, $p = 0.96$; computer and data science: pp change = 1.2%, $t(989) = 0.55$, $p = 0.58$; business: pp change = -0.6%, $t(989) = -0.28$, $p = 0.78$).

## Discussion

This study investigated the role of purpose in academic exploration by encouraging students to reflect on their purpose during course searches. Our findings show that a purpose reflection caused students to be less interested in computer and data science, and more interested in creative arts and social change. The self-interest and self-transcendent purpose interventions produced similar outcomes and primarily affected new users. The intervention encouraged students to be cognitively more engaged during course search but also discouraged students from engaging in search.

One of the main contributions of this study is establishing a causal relationship between purpose reflection and interest exploration. In line with our expectations, students are more interested in areas related to social change when they reflect on their purpose. This may be because making social change reflects an individual's "desire to make a difference in the world" (Damon et al., 2003, p. 121), which is a prominent theme for having a purpose. Students are also more interested in creative arts as a result of purpose reflection. Previous research suggests that creativity can be an important pathway to meaning-making, especially in professional contexts (Kaufman, 2018). When people have more freedom to be creative at work, they find their work more purposeful (Lips-Wiersma & Morris, 2009). Our research provides further support for the importance of creativity in the pursuit of meaning by observing that purpose reflection leads individuals to gravitate toward creative courses. Future empirical research may examine the relationship between purpose and creativity.

We also found that the purpose intervention led to a substantial decrease in students' interest in computer and data science. We did not find a decrease in interest in business. This finding contributes valuable insights into students' motives for considering courses in computer and data science during enrollment. It suggests that students' choices in this area may be primarily driven by motives such as gaining high financial returns in the job market, which are commonly promised by fields of computer science and data science (U.S. & Bureau of Labor Statistics, 2022). In contrast, purpose is associated with motives that are more profound, far-reaching, and high-order (Damon et al., 2003; Pfund et al., 2020). Purpose reflection may have encouraged students to consider more abstract and long-term motives, and thus be less interested in computer and data science. This highlights the need for future research to investigate ways that can help students see more purpose in courses related to computer and data science, especially given the ubiquity of information technology in our daily lives and the potential for technological tools to create social good (Tomašev et al., 2020).

In contrast to our hypotheses, we did not find many significant differences between the self-interest purpose intervention and the self-transcendent purpose intervention in how they affect different areas of interest. Based on the distinction made by Yeager and colleagues (2014), we initially hypothesized that the self-interest purpose intervention may affect students to explore interests that are more self-fulfilling and the self-transcendent purpose intervention may affect them to explore interests that are more related to benefiting others. However, our results showed that both interventions led to similar outcomes, suggesting that students in our sample may not distinguish between the two kinds of purpose. This is consistent with prior research showing that altruistic motives often coexist with egoistic motives (Batson & Powell, 2003; Feiler et al., 2012). Interview research also shows that when students talk about their purpose, they commonly describe their purpose with mixed self-interest and self-transcendent motives (Yeager et al., 2012; Yeager & Bundick, 2009).

**Theoretical Implications**

This research makes theoretical contributions to the literature on purpose. We demonstrate that purpose reflection can be induced through a subtle intervention embedded in a digital tool. Prior research on purpose primarily uses qualitative methods and correlational survey studies (e.g., Bronk, 2012; Burrow & Hill, 2011; Malin et al., 2014), which has limitations in understanding causal relationships between purpose and other psychological or behavioral outcomes (Hill et al., 2013). In recent years, studies have used interventions to experimentally investigate the psychological effects of purpose in educational settings. For example, two studies tested a 20-minute writing assignment to promote purpose during learning among middle school students (Reeves et al., 2021; Yeager et al., 2014). Our study contributes to this research effort by advancing purpose research through randomized field experiments. We demonstrated how a purpose reflection intervention affected course searches among college students. Although we did not measure if students' sense of purpose was raised by the intervention, we observed students being more reflective about their purpose when writing their search query. Compared to previous purpose intervention studies, our intervention is seamlessly integrated into an existing tool and requires no separate activity. We hope this intervention approach may inspire future research to investigate the effect of purpose reflection on various outcomes.

This research also contributes to the literature on interest and motivation. In recent decades, significant accomplishments have been made in understanding and supporting interest development (Hidi & Renninger, 2006; Hulleman et al., 2010). However, empirical studies have primarily focused on supporting interest in a single subject area (e.g., math interest in math classrooms; Kosovich et al., 2019). While this kind of research is insightful for developing targeted approaches to facilitate interest in learning, it falls short of capturing the multiplicity of interests that an individual may have. How an individual reconciles their various interests, especially ones that compete for attention, may affect subsequent interest development. Recent work has called for and attempted to understand interest development under a multiple-interest perspective (e.g., interest across courses and domains; Akkerman & Bakker, 2019; Vulperhorst et al., 2018). The idea of a broader perspective on interest development is also reflected in recent movements to study students' academic progress in higher education, which call for studying shifts in academic interests and academic choices across subject domains to understand how students develop their academic pathways as active agents as opposed to inert entities (Kizilcec et al., 2023). However, empirical and especially experimental studies that investigate multiple interests in authentic academic contexts are rare due to methodological challenges to documenting students' interests across multiple courses (Akkerman & Bakker, 2019; Barron, 2006). Our study suggests that the course selection process is a valuable context to empirically investigate interest development with a multiple-interest perspective. Future research may further investigate how and why certain interests are prioritized and what other factors may influence engagement across interests.

**Practical Implications**

This research has practical implications for institutional leaders and for online course catalog design. We demonstrated a simple and scalable approach that institutions can adopt to help students reflect on their purpose during college experience. Having a sense of purpose is beneficial to students' academic motivation and achievement, subjective well-being, and future orientations (Damon, 2008; Fry, 1998; Sumner et al., 2015), but reflection on purpose is seldom a part of the curriculum except for a few developmental programs offered at some universities. For most students, the closest experience they have with purpose reflection is writing a purpose statement for their college application. However, these purpose statements are rarely brought up again after students arrive on campus. Our study demonstrates that a purpose reflection prompt can influence students' interests. Compared to developmental programs, an online purpose intervention during course search holds two benefits. First, it is a low-cost approach that is integrated into an existing digital platform and can be equitably accessed and scaled to thousands of students at the university. This is substantially less labor-intensive compared to developing and running entire programs. Second, this approach is well integrated with students' authentic learning experiences. By integrating purpose reflection into the course selection process, students can connect their purpose with their course considerations and instantly receive relevant information about course options. This can be considered a constructivist learning experience because it emphasizes learning through contextual, personal experience and active reflection (Bada & Olusegun, 2015; Dewey, 1986). This approach also aligns with prior work emphasizing the significance of the environment and access to timely opportunities for adolescent purpose development (Mariano, 2014). In summary, integrating purpose reflection into course search provides students with the opportunity to actively engage in their own learning and connect their personal experiences with their academic pursuits.

This research also suggests several cautionary considerations with the design of a purpose reflection intervention in online course catalogs. First, we found that the purpose reflection intervention only engaged students once they started writing a search query, but may cause students to opt out of searching in the first place. This may be because the process of purpose reflection, while beneficial, can be unpleasant (Kidwell et al., 1995; Schwartz et al., 2009). Generating a purpose requires insight, introspection, and planning (Carver & Scheier, 2002), which is more cognitively demanding than simply describing one's interest alone. To reduce the cognitive load of purpose reflection, researchers may use common instructional techniques such as scaffolding. In our research, we used worked examples, a kind of scaffolding technique, to facilitate query writing. Many students (40.2%) wrote their search queries according to the example format. Other scaffolding techniques can be tested in the future, such as breaking the purpose reflection prompt down into several smaller questions.

Second, we found that the purpose reflection intervention primarily influenced the interests of new users of the online course catalog, not existing users. One explanation is that existing users have already formed a mental model of the platform and how to use it, which makes it harder to evoke behavioral change even if they notice the new search prompt. This explanation is partially evidenced by our findings on user engagement. We observed that existing users did write more and displayed more causal thinking in the intervention conditions, suggesting that they noticed the search prompt, but their level of cognitive engagement was still lower than new users. Another explanation is that new users pay more attention to the search prompt than existing users. Therefore, future intervention design can consider adding visual cues that can make the changes in the search prompt salient to existing users.

**Limitations**

We discuss five limitations of this study. First, outcomes in this study are limited to observations on an online course catalog. Future research could analyze subsequent course enrollments to examine the downstream impacts of purpose reflection and interest exploration. Second, causal estimates of the relationship between purpose reflection and interest may be biased given the differential disengagement in query writing across conditions. We cannot observe queries written by students who did not search and this selection effect could correlate with students' openness to changing their interests. Higher rates of disengagement are commonly found in opt-in intervention treatments (Kizilcec et al., 2017a; Kizilcec et al., 2017b; Kizilcec et al., 2020). Future work could examine patterns in who tends to disengage as a result of adding a purpose reflection intervention to account for the variation in disengagement during analysis and to design tailored interventions that better engage certain subpopulations.

Third, we did not measure whether the purpose reflection intervention induced a stronger sense of purpose, due to its authentic implementation during course selection. Nevertheless, our results suggest an increased sense of purpose based on students' increased cognitive engagement in query writing. We explicitly avoided using the word "purpose" in the intervention to not create a demand effect. Future studies could include pre- and post-surveys to the intervention to measure psychological constructs and gain better understanding of student experiences that are not directly observable. For example, the intervention may cause an increased sense of purpose along with other psychological constructs, such as meaningfulness. In fact, distinguishing purpose from these constructs has been empirically challenging (Hill et al., 2013).

Fourth, the topic model is limited in capturing the full range of topics in search queries due to its reliance on term co-occurrence. In our context, some terms with low co-occurrence may still belong to the same topic regarding areas of interest. For example, chemistry and physics both belong to the natural sciences but may not be often mentioned together in search queries. Thus, the topic model may categorize them into different topics even though they belong to the same area of interest. While topic models have been used in past intervention research to examine subgroup effects (Kizilcec et al., 2017a), future studies can complement this approach by using supervised learning techniques. For example, researchers can develop topic categories and dictionaries that label terms in each category in a training dataset of search queries. Prediction models can then be used to assign search queries to topic categories in a full dataset.

Fifth, the consequences of a purpose reflection intervention may be undesirable in contexts that prioritize academic progress on linear pathways to achieve timely graduation for a population of students who are at risk of dropping out of college (Jenkins & Cho, 2013). In these contexts, a purpose reflection at the beginning of a student's academic pathway may be most appropriate and beneficial to encourage deliberate academic choices aligned with the student's purpose. Future work should evaluate the impact of this type of intervention in institutional contexts with more structured curriculums and lower graduation rates.

## Conclusion

Online course catalogs play an important role in college students' interest exploration and course consideration. In this study, we examined a subtle purpose reflection intervention in an online course catalog across over 4,000 college students through one whole academic year. Our findings show that the intervention caused students to be more engaged in search query writing, but less engaged in search activities. The intervention also led students to explore the areas of social change and creative arts over computer and data science. Overall, this study makes novel contributions to understanding how interest exploration can be influenced during course consideration. The study also has practical implications that demonstrate a scalable approach for purpose reflection using an online course catalog as a digital medium.

## Acknowledgments

We thank Fynn Datoo for his technical support. This work was supported in part by Cornell's Vice Provost Office for Academic Innovation and Office of University Registrar.

## Appendix

### A. Bigram Tokenization

In search queries, bigrams often can be more indicative of topics than unigrams. For example, although "learning sciences" and "machine learning" both include the word "learning", they refer to completely different disciplinary areas in which a bigram tokenization can differentiate but a unigram tokenization cannot. The addition of bigrams is especially useful for topic modeling given that search queries are "short documents". Bigrams can provide more information to infer topics in documents. The ten most frequent bigrams that we obtained from students' search queries are: "machine learning" (2.0% of all searches), "computer science" (1.5%), "data science" (1.4%), "artificial intelligence" (1.0%), "arts sciences" (1.0%), "real estate" (0.8%), "climate change" (0.7%), "information science" (0.5%), "data analytics" (0.5%), "classics class" (0.5%). We added bigrams to the list of tokens together with their unigrams as opposed to replacing the corresponding unigrams. For example, we added the bigram 'machine learning' together with the unigrams 'machine' and 'learning' to the document they belong to.

We opted for this decision for two reasons. First, we want to remain agnostic as to whether it is the bigrams or the unigrams that contributed to the meaning of the words, given the low proportion of even the most frequent bigrams. Replacing bigrams with their corresponding unigrams assumes that the unigrams are a false interpretation of the real meaning of the words. Adding bigrams with their corresponding unigrams assumes that both unigrams and bigrams contribute equally to the real meaning of the words. Second, previous work suggested that replacing bigrams in documents to account for them in the topic model only mildly improves or does not impact the performance of the topic model (Lau et al 2013). In our study, we found that the interpretability of the topic modeling results became better when we added bigrams instead of replacing unigrams with bigrams.

**B. Determining the Optimal Number of Topics in Topic Modeling**

To determine the optimal number of topics in topic modeling, we first set the range of topic numbers to be 2 to 10. Thus, the granularity of topics would satisfy our research goal, which is to find broad subject areas, such as engineering, humanities, natural science, and business. Then, we used the UMass metric to evaluate within-topic coherence (Roder, et al., 2015). The closer the UMass score is to zero, the more coherent the topics are. We systematically examined the topic modeling results across 2 to 10 topics and across 5 to 15 minimum bigram frequencies. For every combination of parameters (i.e., topics and minimum bigram frequencies), we ran the model 20 times to obtain a robust evaluation of the coherence score. For every topic modeling result, we carefully discussed the interpretability of the results based on the most frequent tokens in the topics and the 20 most representative students in each topic based on the topic probabilities. Specifically, each author first independently evaluated and ranked the results based on interpretability. Then, we compared and discussed the ranking together. We found that changing the minimum bigram frequency did not affect the interpretability of the topic modeling results much once the value was above 10. We also found that when the number of topics is below 5, there are always topics that are mixed with terms from several distinctive subject areas that make it difficult to come up with a topic label. When comparing the number of topics from 6 to 10, all authors ranked eight topics as the most optimal number of topics in terms of within-topic coherence, interpretability, and meaningfulness for the research purpose. In the end, we decided on using eight topics with a minimum bigram frequency of ten as the topic modeling results for further analysis. The authors together labeled the topics based on the most frequent terms (i.e., unigrams and bigrams) in each topic and the search queries with the highest scores in each topic (Table B1). We relied more heavily on representative search queries because some frequent terms appear commonly across more than one topic and the most representative search queries provide important contextual information on the topic.

    After topic labeling, we merged the eight topics into five areas of interest to fit for our research purpose. First, we found that not all topics were on the same level of granularity. Specifically, Topic 8 "Finance" is a subset of Topic 4 "Business". Therefore, we combined Topic 8 "Finance" and Topic 4 "Business" into "Business". Second, we found that some topics are closely related. Specifically, Topic 2 and Topic 7 are both related to creative arts, and thus we merged them. Third, we found that Topic 3 "Human and the Universe" is too broad to be defined as a coherent area, and Topic 6 "Biology" may be too specific to be counted as an area for our study. In the end, we obtained five areas of interest (Table B2).

**Table B1**

*The Eight Topics Produced by the Topic Model*

| Topic | Avg. Probability | Most Frequent Terms | Example Query | Topic Label |
|---|---|---|---|---|
| 1 | 18.2% | Data, science, learning, machine, machine_learning, computer, data_science, theory, intelligence, design, real | *"I want to learn about virtual reality machine learning and computer science"* | Computer and Data Science |
| 2 | 14.5% | Art, design, history, writing, music, creative, science, food, business, sustainability, wine | *"I want to understand how to get talents to thinking creatively to give an organization a competitive advantage"* | Creative Arts |
| 3 | 14.4% | Health, psychology, science, liberal, studies, quantum, physics, social, human, mental, astronomy | *"I want to learn about applied physics and atmospheric physics"* | Humans and the Universe |
| 4 | 13.0% | Business, law, management, marketing, consulting, design, environmental, human, development, social, technology | *"I want to learn about the business side of the biotechnology industry including venture capital and management"* | Business |
| 5 | 12.1% | Climate, animals, world, change, science, climate_change, law, politics, philosophy, social, music | *"I want to learn about feminism and misogyny across the world and how gender roles have changed over time"* | Social Change |
| 6 | 9.6% | Biology, medicine, human, art, neuroscience, engineering, biomg*, psychology, anatomy, cooking, chemistry | *"I want to learn about cellular immunology and genetics for research purposes"* | Biology |
| 7 | 8.4% | Arts, history, sciences, arts_sciences, latin, design, natural, research, field, writing, birds | *"I want to engage in interdisciplinary research on the relationship between people landscapes and histories"* | Creative arts |
| 8 | 9.0% | Finance, leadership, analytics, business, technology, writing, data, data_analytics, financial, management, tech | *"I want to learn about finance and banking because I want to go into banking"* | Finance |

\* *biomg* stands for Molecular Biology and Genetics, an acronym of a subject code at the present university.

**Table B2**

*The Five Areas of Interest Derived from the Eight Topics*

| Topics Combined | Average Probability | Label for Areas of Interest |
|---|---|---|
| Topic 4, Topic 8 | 22.0% | Business |
| Topic 1 | 18.2% | Computer and Data Science |
| Topic 2, Topic 7 | 22.9% | Creative Arts |
| Topic 5 | 12.1% | Social Change |
| Topic 3, Topic 6 | 24% | Others |

INTEREST EXPLORATION IN AN ONLINE COURSE CATALOG											41Harrison, M. H., Hernandez, P. A., & Stevens, M. L. (2022). Should I start at math 101? Content repetition as an academic strategy in elective curriculums. *Sociology of Education*, *95*(2), 133–152.

Hidi, S., & Harackiewicz, J. M. (2000). Motivating the academically unmotivated: A critical issue for the 21st century. *Review of Educational Research*, *70*(2), 151–179.

Hidi, S., & Renninger, K. A. (2006). The four-phase model of interest development. *Educational Psychologist*, *41*(2), 111–127.

Hill, P. L., Burrow, A. L., & Sumner, R. (2013). Addressing important questions in the field of adolescent purpose. *Child Development Perspectives*, *7*(4), 232–236.

Hinds, E. M., & Shultz, G. V. (2018). Investigation of the factors that influence undergraduate student chemistry course selection. *Journal of Chemical Education*, *95*(6), 913–919.

Hong, L., & Davison, B. D. (2010). Empirical study of topic modeling in twitter. *Proceedings of the First Workshop on Social Media Analytics*, 80–88.

Hulleman, C. S., Godes, O., Hendricks, B. L., & Harackiewicz, J. M. (2010). Enhancing interest and performance with a utility value intervention. *Journal of Educational Psychology*, *102*(4), 880.

Hulleman, C. S., & Harackiewicz, J. M. (2009). Promoting interest and performance in high school science classes. *Science*, *326*(5958), 1410–1412.

Jacobs, T., & Tschötschel, R. (2019). Topic models meet discourse analysis: a quantitative tool for a qualitative approach. *International Journal of Social Research Methodology*, *22*(5), 469–485.

Jenkins, D., & Cho, S. W. (2013). Get with the program… and finish it: Building guided pathways to accelerate student completion. *New Directions for Community Colleges, 2013*(164), 27-35.

Joksimovic, S., Gasevic, D., Kovanovic, V., Adesope, O., & Hatala, M. (2014). Psychological characteristics in cognitive presence of communities of inquiry: A linguistic analysis of online discussions. *The Internet and Higher Education*, *22*, 1–10.

Jordaan, J. P. (1963). Exploratory behavior: The formation of self and occupational concepts. *Career Development: Self-Concept Theory*, 42-78.

INTEREST EXPLORATION IN AN ONLINE COURSE CATALOG                                44